\documentstyle[12pt]{article}
\newcommand{\nc}{\newcommand}
\nc{\renc}{\renewcommand}

\nc{\etal}{\mbox{\it et al. }}
\nc{\ie}{{\it i.e.}}
\nc{\eg}{{\it e.g.}}

\renc{\thefootnote}{\arabic{footnote}}
\nc{\capt}[1]{{\bf Figure.} {\small\sl #1}}

\nc{\eqs}[2]{\mbox{Eqs.~(\ref{#1},\,\ref{#2})}}
\nc{\eq}[1]{\mbox{Eq.~(\ref{#1})}}

\nc{\figs}[2]{\mbox{Figs.~(\ref{#1},\,\ref{#2})}}
\nc{\fig}[1]{\mbox{Fig~.(\ref{#1})}}

\nc{\tag}[1]{\label{#1} \marginpar{{\footnotesize #1}}}
\nc{\mtag}[1]{\label{#1} \mbox{\marginpar{{\footnotesize #1}}}}
\renc{\baselinestretch}{1.2}
\newlength{\overeqskip}
\newlength{\undereqskip}
\nc{\be}[1]{\begin{equation} \mbox{$\label{#1}$}}
\nc{\bea}[1]{\begin{eqnarray} \mbox{$\label{#1}$}}
\nc{\Section}[2]{\section{#2}\label{#1}}
\nc{\Bibitem}[1]{\bibitem{#1}}
\nc{\Label}[1]{\label{#1}}

\nc{\eea}{\vspace{\undereqskip}\end{eqnarray}}
\nc{\ee}{\vspace{\undereqskip}\end{equation}}
\nc{\bdm}{\begin{displaymath}}
\nc{\edm}{\end{displaymath}}
\nc{\dpsty}{\displaystyle}
\nc{\bc}{\begin{center}}
\nc{\ec}{\end{center}}
\nc{\ba}{\begin{array}}
\nc{\ea}{\end{array}}
\nc{\bab}{\begin{abstract}}
\nc{\eab}{\end{abstract}}
\nc{\btab}{\begin{tabular}}
\nc{\etab}{\end{tabular}}
\nc{\bit}{\begin{itemize}}
\nc{\eit}{\end{itemize}}
\nc{\ben}{\begin{enumerate}}
\nc{\een}{\end{enumerate}}
\nc{\bfig}{\begin{figure}}
\nc{\efig}{\end{figure}}
\nc{\figcap}[1]{\begin{quote}\refstepcounter{figure}
        {\bf Figure \thefigure}: {\small #1}\end{quote}}
\nc{\arreq}{&\!=\!&}
\nc{\arrmi}{&\!-\!&}
\nc{\arrpl}{&\!+\!&}
\nc{\arrap}{&\!\!\!\approx\!\!\!&}
\nc{\non}{\nonumber\\*}
\nc{\align}{\!\!\!\!\!\!\!\!&&}

\def\lsim{\; \raise0.3ex\hbox{$<$\kern-0.75em
      \raise-1.1ex\hbox{$\sim$}}\; }
\def\gsim{\; \raise0.3ex\hbox{$>$\kern-0.75em
      \raise-1.1ex\hbox{$\sim$}}\; }
\nc{\DOT}{\hspace{-0.08in}{\bf .}\hspace{0.1in}}
\nc{\Laada}{\hbox {$\sqcap$ \kern -1em $\sqcup$}}
\nc\loota{{\scriptstyle\sqcap\kern-0.55em\hbox{$\scriptstyle\sqcup$}}}
\nc\Loota{{\sqcap\kern-0.65em\hbox{$\sqcup$}}}
\nc\laada{\Loota}
\nc{\qed}{\hskip 3em \hbox{\BOX} \vskip 2ex}

\nc{\real}{{\rm I \! R}}
\nc{\Z}{{\sf Z \!\!\! Z}}
\nc{\complex}{{\rm C\!\!\! {\sf I}\,\,}}
\def\bigid{\leavevmode\hbox{\small1\kern-3.8pt\normalsize1}}
\def\id{\leavevmode\hbox{\small1\kern-3.3pt\normalsize1}}
\nc{\slask}{\!\!\!/}
\nc{\bis}{{\prime\prime}}
\nc{\pa}{\partial}
\nc{\na}{\nabla}
\nc{\ra}{\rangle}
\nc{\la}{\langle}
\nc{\goto}{\rightarrow}
\nc{\swap}{\leftrightarrow}

\nc{\EE}[1]{ \mbox{$\cdot10^{#1}$} }
\nc{\abs}[1]{\left|#1\right|}
\nc{\at}[2]{\left.#1\right|_{#2}}
\nc{\norm}[1]{\|#1\|}
\nc{\abscut}[2]{\Abs{#1}_{\scriptscriptstyle#2}}
\nc{\vek}[1]{{\rm\bf #1}}
\nc{\integral}[2]{\int\limits_{#1}^{#2}}
\nc{\inv}[1]{\frac{1}{#1}}
\nc{\dd}[2]{{{\partial #1}\over{\partial #2}}}
\nc{\ddd}[2]{{{{\partial}^2 #1}\over{\partial {#2}^2}}}
\nc{\dddd}[3]{{{{\partial}^2 #1}\over
	{\partial #2 \partial #3}}}
\nc{\dder}[2]{{{d #1}\over{d #2}}}
\nc{\ddder}[2]{{{d^2 #1}\over{d {#2}^2}}}
\nc{\dddder}[3]{{d^2 #1}\over
	{d #2 d #3}}
\nc{\dx}[1]{d\,^{#1}x}
\nc{\dy}[1]{d\,^{#1}y}
\nc{\dz}[1]{d\,^{#1}z}
\nc{\dl}[1]{\frac{d\,^{#1}l}{(2\pi)^{#1}}}
\nc{\dk}[1]{\frac{d\,^{#1}k}{(2\pi)^{#1}}}
\nc{\dq}[1]{\frac{d\,^{#1}q}{(2\pi)^{#1}}}

\nc{\cc}{\mbox{$c.c.$ }}
\nc{\hc}{\mbox{$h.c.$ }}
\nc{\cf}{cf.\ }
\nc{\erfc}{{\rm erfc}}
\nc{\Tr}{{\rm Tr\,}}
\nc{\tr}{{\rm tr\,}}
\nc{\pol}{{\rm pol}}
\nc{\sign}{{\rm sign}}
\nc{\bfT}{{\bf T }}

\nc{\cA}{{\cal A}}
\nc{\cB}{{\cal B}}
\nc{\cD}{{\cal D}}
\nc{\cE}{{\cal E}}
\nc{\cG}{{\cal G}}
\nc{\cH}{{\cal H}}
\nc{\cL}{{\cal L}}
\nc{\cO}{{\cal O}}
\nc{\cT}{{\cal T}}
\nc{\cN}{{\cal N}}
\nc{\rvac}[1]{|{\cal O}#1\rangle}
\nc{\lvac}[1]{\langle{\cal O}#1|}
\nc{\rvacb}[1]{|{\cal O}_\beta #1\rangle}
\nc{\lvacb}[1]{\langle{\cal O}_\beta #1 |}
\nc{\bb}{\bar{\beta}}
\nc{\bt}{\tilde{\beta}}
\nc{\ctH}{\tilde{\cal H}}
\nc{\chH}{\hat{\cal H}}
\nc{\1}{\aa}
\nc{\2}{\"{a}}
\nc{\3}{\"{o}}
\nc{\4}{\AA}
\nc{\5}{\"{A}}
\nc{\6}{\"{O}}
\nc{\al}{\alpha}
\nc{\g}{\gamma}
\nc{\Del}{\Delta}
\nc{\e}{\epsilon}
\nc{\eps}{\epsilon}
\nc{\lam}{\lambda}
\nc{\om}{\omega}
\nc{\Om}{\Omega}
\nc{\ve}{\varepsilon}
\nc{\mn}{{\mu\nu}}
\nc{\k}{\kappa}
\nc{\vp}{\varphi}

\nc{\advp}[3]{{\it  Adv.\ in\ Phys.\ }{{\bf #1} {(#2)} {#3}}}
\nc{\annp}[3]{{\it  Ann.\ Phys.\ (N.Y.)\ }{{\bf #1} {(#2)} {#3}}}
\nc{\apl}[3]{{\it  Appl. Phys. Lett. }{{\bf #1} {(#2)} {#3}}}
\nc{\apj}[3]{{\it  Ap.\ J.\ }{{\bf #1} {(#2)} {#3}}}
\nc{\apjl}[3]{{\it  Ap.\ J.\ Lett.\ }{{\bf #1} {(#2)} {#3}}}
\nc{\app}[3]{{\it Astropart.\ Phys.\ }{{\bf #1} {(#2)} {#3}}}  
\nc{\cmp}[3]{{\it  Comm.\ Math.\ Phys.\ }{{ \bf #1} {(#2)} {#3}}}
\nc{\cqg}[3]{{\it  Class.\ Quant.\ Grav.\ }{{\bf #1} {(#2)} {#3}}}
\nc{\epl}[3]{{\it  Europhys.\ Lett.\ }{{\bf #1} {(#2)} {#3}}}
\nc{\ijmp}[3]{{\it Int.\ J.\ Mod.\ Phys.\ }{{\bf #1} {(#2)} {#3}}}
\nc{\ijtp}[3]{{\it Int.\ J.\ Theor.\ Phys.\ }{{\bf #1} {(#2)} {#3}}}
\nc{\jmp}[3]{{\it  J.\ Math.\ Phys.\ }{{ \bf #1} {(#2)} {#3}}}
\nc{\jpa}[3]{{\it  J.\ Phys.\ A\ }{{\bf #1} {(#2)} {#3}}}
\nc{\jpc}[3]{{\it  J.\ Phys.\ C\ }{{\bf #1} {(#2)} {#3}}}
\nc{\jap}[3]{{\it J.\ Appl.\ Phys.\ }{{\bf #1} {(#2)} {#3}}}
\nc{\jpsj}[3]{{\it J.\ Phys.\ Soc.\ Japan\ }{{\bf #1} {(#2)} {#3}}}
\nc{\lmp}[3]{{\it Lett.\ Math.\ Phys.\ }{{\bf #1} {(#2)} {#3}}}
\nc{\mpl}[3]{{\it  Mod.\ Phys.\ Lett.\ }{{\bf #1} {(#2)} {#3}}}
\nc{\ncim}[3]{{\it  Nuov.\ Cim.\ }{{\bf #1} {(#2)} {#3}}}
\nc{\np}[3]{{\it  Nucl.\ Phys.\ }{{\bf #1} {(#2)} {#3}}}
\nc{\pr}[3]{{\it Phys.\ Rev.\ }{{\bf #1} {(#2)} {#3}}}
\nc{\pra}[3]{{\it  Phys.\ Rev.\ A\ }{{\bf #1} {(#2)} {#3}}}
\nc{\prb}[3]{{\it  Phys.\ Rev.\ B\ }{{{\bf #1} {(#2)} {#3}}}}
\nc{\prc}[3]{{\it  Phys.\ Rev.\ C\ }{{\bf #1} {(#2)} {#3}}}
\nc{\prd}[3]{{\it  Phys.\ Rev.\ D\ }{{\bf #1} {(#2)} {#3}}}
\nc{\prl}[3]{{\it Phys\ Rev.\ Lett.\ }{{\bf #1} {(#2)} {#3}}}
\nc{\pl}[3]{{\it  Phys.\ Lett.\ }{{\bf #1} {(#2)} {#3}}}
\nc{\prep}[3]{{\it Phys\. Rep.\ }{{\bf #1} {(#2)} {#3}}}
\nc{\prsl}[3]{{\it Proc.\ R.\ Soc.\ London\ }{{\bf #1} {(#2)} {#3}}}
\nc{\ptp}[3]{{\it  Prog.\ Theor.\ Phys.\ }{{\bf #1} {(#2)} {#3}}}
\nc{\ptps}[3]{{\it  Prog\ Theor.\ Phys.\ suppl.\ }{{\bf #1} {(#2)} {#3}}}
\nc{\physa}[3]{{\it  Physica\ A\ }{{\bf #1} {(#2)} {#3}}}
\nc{\physb}[3]{{\it  Physica\ B\ }{{\bf #1} {(#2)} {#3}}}
\nc{\phys}[3]{{\it Physica\ }{{\bf #1} {(#2)} {#3}}}
\nc{\rmp}[3]{{\it  Rev.\ Mod.\ Phys.\ }{{\bf #1} {(#2)} {#3}}}
\nc{\rpp}[3]{{\it Rep.\ Prog.\ Phys.\ }{{\bf #1} {(#2)} {#3}}}
\nc{\sjnp}[3]{{\it Sov.\ J.\ Nucl.\ Phys.\ }{{\bf #1} {(#2)} {#3}}}
\nc{\spjetp}[3]{{\it Sov.\ Phys.\ JETP\ }{{\bf #1} {(#2)} {#3}}}
\nc{\yf}[3]{{\it Yad.\ Fiz.\ }{{\bf #1} {(#2)} {#3}}}
\nc{\zetp}[3]{{\it Zh.\ Eksp.\ Teor.\ Fiz.\  }{{\bf #1}  {(#2)} {#3}}}
\nc{\zp}[3]{{\it Z.\ Phys.\ }{{\bf #1} {(#2)} {#3}}}
\nc{\ibid}[3]{{\sl ibid.\ }{{\bf #1} {#2} {#3}}}
\nc{\rf}[1]{(\ref{#1})}
\nc{\nn}{\nonumber \\*}
\nc{\bfB}{\bf{B}}
\nc{\bfv}{\bf{v}}
\nc{\bfx}{\bf{x}}
\nc{\bfy}{\bf{y}}
\nc{\vx}{\vec{x}}
\nc{\vy}{\vec{y}}
\nc{\oB}{\overline{B}}
\nc{\oI}{\overline{I}}
\nc{\oR}{\overline{R}}
\nc{\rar}{\rightarrow}
\nc{\ti}{\times}
\nc{\slsh}{\hskip-5pt/}
\nc{\sm}{Standard~Model~}
\nc{\MP}{M_{\rm Pl}}
\nc{\tp}{t_{\rm Pl}}
\nc{\ave}{\bar{E}}

\renc{\min}{p_{\rm min}}
\renc{\max}{p_{\rm max}}
\nc{\pmin}{p_{\rm min}}
\nc{\pmax}{p_{\rm max}}
\nc{\fo}{f_0}
\nc{\foi}{f_{0,i}\,}
\nc{\fop}{f_0^P}
\nc{\fou}{f_0^U}
\def\sepand{\rule{14cm}{0pt}\and}
\nc{\eff}{{\rm eff}}
\nc{\MT}{M_{\rm T}}
\nc{\ML}{M_{\rm L}}
\nc{\kk}{\vek{k}}
\nc{\pp}{{\rm p}}
\nc{\cb}{critical bubble~}
\nc{\cbs}{critical bubbles~}
\nc{\scb}{subcritical bubble~}
\nc{\scbs}{subcritical bubbles~}
\nc{\bh}{{BH}}
\nc{\BH}{black hole\ }
\nc{\BHs}{black holes\ }
\nc{\mbh}{m_{BH}}
\begin{document}

{\title{{\hfill {{\small  
        }}\vskip 1truecm}
{\bf Non-equilibrium universe and \BH evaporation}}

 
\author{
{\sc Iiro Vilja$^{1}$ }\\ 
{\sl Department of Physics,
University of Turku} \\
{\sl FIN-20014 Turku, Finland} \\
\sepand
}
\date{ }
\maketitle}
\vspace{2cm}
\begin{abstract}
\noindent The evaporation of the \BHs during the very early universe is 
studied. Starting from \BH filled universe, the distributions of particle
species are calculated and showed, that they differ remarkably from the 
corresponding equilibrium distribution. This may have great impact to the
physics of the very early universe. Also the evolution of the universe 
during the evaporation has been studied.
\end{abstract}
\vfill
\footnoterule
{$^1$vilja@newton.tfy.utu.fi}
\thispagestyle{empty}
\newpage
\setcounter{page}{1}

It is well known that in the very early universe the particles
could not been equilibrated due to a too short equilibration time
\cite{1,ES,EEV1}. Indeed, the only possibility to have equilibrium
distributions at times $t \sim 10^{-34}$ s corresponding equilibration 
temperature $T_{eq} \sim 10^{15}$ K is to assume, that there is some 
process producing particles readily to equilibrium distribution. (It
should be noted that in this case it is not really question of
thermodynamical equilibrium because there are no interactions
maintaining it.) Sometimes it is assumed that {\it e.g.} black hole 
evaporation could be such a process \cite{EEV1}. When the particles 
are originated from small black holes near Planck scale which evaporate due 
to Hawking process\cite{haw}, the radiation at any given time 
is thermal with temperature equal to Hawking temperature. 

In the present paper we calculate the distribution of particles originated
from black holes. We suppose that the black holes are generated at Planck
time through quantum gravitation. After Planck time they begin to decay 
emitting particles. We show that the resulting particle distributions are not
the equilibrium one, but differs crucially from it. This difference may play
remarkable role {\it e.g.} in inflation \cite{infl}\footnote{The existence 
of small \BHs at early times would have also many other interesting
consequences, see \cite{plen}.}, because the 
effective, ensemble corrected potential \cite{EEV1} differs now from the one 
calculated with an equilibrium thermal bath \cite{RV}.

In the present study, for simplicity, we neglect the curvature effects,
arising at Planck scale due to ambiguous concept of vacuum state \cite{BD}.
By assuming isotropy, the comoving observers have same vacuum and hence,
whenever the average energy per particle is clearly smaller than Planck mass,
the analysis should be applicable. Consistently we omit also all direct
couplings of quantum fields to curvature and treat the gravitational field
classically.

Let the evolution of the universe be given by the scale parameter $R(t)$ and
suppose that at Planck time no other matter exists in the universe than 
black holes. The radius of a (non-rotating) black hole with mass $m_\bh$ 
is given by $r_\bh = 2m_\bh /\MP^2$, where $\MP$ is the Planck mass. The \BH 
surface area is now given by
\be{area}
A_\bh = {16\pi m_\bh^2\over\MP^4}.
\ee
For simplicity,
suppose that all \BHs have same mass $m_i$  at initial time and their number 
density $t_i$ is $n(t_i)$. Because there is no other type of matter than 
black holes, the universe is at $t_i$ matter dominated, {\it i.e.} 
$R(t) \propto t^{2/3}$. The ratio of the density of the universe to the 
critical density at that time is now given by 
$\Omega (t_i) = 8 \pi m_i n(t_i) /(3 \MP^2 H(t_i)^2)$, where the 
Hubble parameter is $H(t_i) = \frac 23 t_i^{-1}$. Thus we can parametrize the
initial density with density ratio $\Omega_i$. 

Because a \BH  radiates at any time thermal radiation with the temperature
given by the Hawking temperature $T_\bh = \MP^2/(8\pi m_\bh )$ through 
its surface, the loss of energy in time unit is $A_\bh h_* T_\bh^4 \pi^2/30$,
where $h_*$ is an effective number of particles lighter than \BH mass. So, as
well known, the dynamics of a \BH evaporation is governed by the equation
\be{BHD}
\dot m_\bh = - {h_*\over 15\times 8^3\pi^2} {\MP^4\over\mbh^2},
\ee
where we have omitted the absorption of radiation back to the black holes. This 
is well motived whenever the radiation density $\cE_{rad}$ is small, {\it i.e.}
$\cE_{rad} A_\bh \ll |\dot\mbh |$, which is the case. The solution of 
the equation \rf{BHD} reads 
\be{sBHD}
\mbh (t) = m_i \left ( 1 - { t - t_i\over \tau}\right )^{1/3},
\ee
where the \BH life-time $\tau $ is given by
\be{tau}
\tau = {5\times 8^3 \pi^2 m_i^3 \over \MP^4 h_*}.
\ee
We express the initial \BH mass in terms of the inverse of $\tau$
\be{mi}
m_i = {\MP\over 8} \left ({h_*\over 5\pi^2\delta}\right )^{1/3},
\ee
where $\delta \equiv 1/(\MP \tau)$ has to be a small number $ \ll 1$, 
{\it i.e.} the
\BH life-time is much longer than Planck time\footnote{Indeed,
one has to assume that $t_i + \tau \gg t_{Pl}$, because otherwise
quantum gravitational effects would certainly not be omitted.}. Using the
expression for $\Omega_i$ and \rf{mi} we can give the \BH number density
\be{n}
n(t_i) = \frac 43 {\MP\over t_i^2} \left ({5\over\pi h_*}\right )^{1/3}
\Omega_i \delta^{1/3}.
\ee

We are now ready to calculate the distribution of the particles emitted by 
the black holes. We further simplify the situation by stating that all emitting
particles are essentially massless, so that the effective particle number
$h_*$ remains constant during the evaporation\footnote{The massive
case is as well calculable as the massless one, but leads to introducing of 
numerous new parameters, which is avoided here.}. Strictly speaking, this is 
not true, but gives a reasonable approximation of the situation, where no 
specific model with its particle contents is given. Also, we suppose that the \BHs decay before the thermalization starts. This means that no thermalizating 
processes are effective during the evaporation and hence not taken into account in the calculation. Whether this is a good assumption or not is studied later
in this paper.

During a time $[t,\ t + dt]$ ($t > t_i$) the energy emitted by a \BH 
and distributed to $h_*$ particle species (degrees of freedom) is
$dE = - \dot \mbh (t)\, dt$. Because at a given time the radiation emitted 
obeys thermal spectrum with temperature $T_\bh$ ($\beta_\bh = 1/T_\bh$), the 
energy per degree of freedom within momentum between $p$ and $p + dp$ is 
given by
\be{apu1}
dE(p+dp) - dE(p) = {k_\pm dE\over h_*}{\omega (p) f^\pm (p, \beta_\bh(t))d^3p\over
\int d^3k\, \omega (k) f^\pm (k, \beta_\bh(t))},
\ee
where the fraction $k_\pm dE/h_*$ represents the (total) energy per degree of 
freedom ($k_\pm$ is 1 for bosons and $\frac 78$ for fermions appearing due to 
different statistics). The other fraction expresses the part of the energy 
within the given momentum range and the functions $f^\pm$ are the usual 
Fermi-Dirac (+) and Bose-Einstein (-) distribution functions. Thus, the 
particle number density (at the given momentum range) is simply given by
\be{apu2}
{dE(p+dp) - dE(p)\over \omega (p)} = 
{k_\pm dE\over h_*}{f^\pm (\beta_\bh p)\over
\int d^3k\, \omega (k) f^\pm (\beta_\bh k)}.
\ee
Taking into account that the \BH density reads 
$n(t) = n(t_i) \left (R(t_i)/R(t)\right )^3$, one obtains the distribution 
$df_{f(b)}$ of a degree of freedom:
\be{apu3}
df_{f(b)}(p, t) {d^3p\over(2\pi)^3} = n(t) {dE(p+dp) - dE(p)\over \omega (p)}.
\ee
Combining \rf{apu2} and \rf{apu3} we write
\be{ddis}
df_{f(b)} = - n(t) {k_\pm \dot\mbh (t)\, dt\over h_*}{(2\pi)^3 
f^\pm (\beta_\bh p)\over
\int d^3k\, \omega (k) f^\pm (\beta_\bh k)}.
\ee

Eq. \rf{ddis} represents the distribution generated during a time-interval
$[t,\ t+dt]$. The overall distribution thus if we want to end up to at a given 
time $t$ is obtained by summing all contribution from $t_i$ to $t$, {\it i.e.} 
all times $t', \ \ t_i \leq t' \leq t$. These contribution 
are, however, red-shifted by a factor $R(t')/R(t)$, hence the final
particle distribution at time $t$ is given by
\be{dis}
f_{f(b)}(p;t) = - \int_{t_i}^t dt' n(t') {k_\pm \dot\mbh (t')\over h_*}
{(2\pi)^3 f^\pm ({R(t)\over R(t')}\beta_\bh(t') p)\over
\int d^3k\, \omega (k) f^\pm (\beta_\bh(t') k)}.
\ee
The formula above, Eq. \rf{dis}, describes in principle completely
the particle distributions created by \BH evaporation. It can easily be 
generalized to the case, where the (initial) \BHs are not equally massive,
but have some distribution. In this case one just have to sum over all 
contributions emerging due to different \BH masses.

To be able to calculate the distributions in practise, one has to solve the 
evolution of the scale parameter $R(t)$. For that purpose we have to determine
first the time $t_{EQ}$ when the universe enters the radiation dominated
era after the beginning of the \BH evaporation. We have to calculate
when the energy density of the black holes, $\cE_{BH}$, equals the energy
density of radiation,
$\cE_{rad}$, they have emitted: $\cE_{BH}(t_{EQ}) = \cE_{rad}(t_{EQ})$.

Now, the (average) energy density of the black holes at a given time $t$
is given by $\cE_{BH}(t) = \mbh (t) n(t_i) (R(t_i)/R(t))^3$.
During a matter dominated time $R(t) \propto t^{2/3}$ and thus
\be{emat}
\cE_{BH}(t) = \mbh (t) n(t_i)\left ({t_i\over t}\right )^2,\ \ \ \ t_i<t<t_{EQ}.
\ee
On the other hand the radiation energy density at $t < t_{EQ}$ is given by
\be{erad}
\cE_{rad}(t) = \sum_i g_i \int \dk 3 f_i (k, t) \omega_i(k) = 
g_f\int \dk 3 f_f (k, t)\omega(k) + g_b\int \dk 3 f_b (k, t)\omega(k),
\ee
where we have summed over all relevant bosonic ($g_b$) and fermionic ($g_f$)
degrees of freedom\footnote{Here we remind that, for sake of simplicity,
we have assumed that all relevant particles are essentially massless, thus
$\omega(k) = k$.}. Eq. \rf{erad} can simplyfied to
\be{erad2}
\cE_{rad} = -\int_{t_i}^t dt' \dot \mbh (t') \left ({R(t')\over R(t)}\right )^4
n(t_i) \left ({R(t_i)\over R(t')}\right )^3,
\ee
where $- \dot \mbh (t') \left ({R(t')\over R(t)}\right )^4$ corresponds
the energy radiated at $t'$ by \BHs and red-shifted until time $t$, whereas
$n(t_i) \left ({R(t_i)\over R(t')}\right )^3$ represents the number density of 
decaying \BHs at the time when the radiation was produced. 
Evidently, Eq. \rf{erad} can be obtained directly without using any
knowledge about particle distributions. Indeed, the connection between
Eqs. \rf{erad} and \rf{erad2} requires the identification $h_* = g_*$, where
$g_*$ is the usual effective number of degrees of freedom
$g_* = g_b + \frac 78 g_f$. In the rest of the paper we apply this 
identification.

It should be noted here, that the correct evolution of the cosmic scale parameter at matter dominated era is not really as simple as
$R(t) \propto t^{2/3}$ but more complicated. This is due to the fact that 
the correct energy density of matter, {\it i.e.} \BHs at that time reads 
\be{corr}
\cE_{BH} (t) = m(t) n(t_i) (R(t_i)/R(t))^3,
\ee 
which is not as simple as just dilution of the matter.Applying this equation 
to the Einstein equation, one obtains 
\be{realmatt}
R(t) =\frac 32 \left ({8\pi \cE (t_i)\over 3 \MP^2}\right )^{1/2}
{6\tau\over 7} \left [ 1 - \left ({m(t)\over m_i}\right )^{7/2}\right ]^{2/3}
\ee
which, however, is in quite good agreement with the usual matter dominated 
form. Moreover, the calculation 
of the scale parameter at the radiation dominated era leads to a complicated 
integro-differential equation. Therefore we choose here to use the conventional
forms of the evolution parameter $R$, which anyway gives good impression of
the physics involved.
It is now simple matter to take the expression for $\cE_{BH}$ and $\cE_{rad}$
from Eqs. \rf{emat} and \rf{erad2} (and relation $R(t) \propto t^{2/3}$)
and equate them. One obtains a rather simply equation
\be{teq}
\mbh (t_{EQ}) t_{EQ}^{2/3} = - \int_{t_i}^{t_{EQ}} dt\, t^{2/3} \dot\mbh (t).
\ee
It is noteworthy, that eq. \rf{teq} does not depend on the initial \BH 
number density $n(t_i)$ but only on initial time and \BH mass. 
According to the approximated analysis, the radiation 
dominance begins quite late compared to the \BH life-time. When $t_i/\tau$
varies from 0 to 0.5, the combination $(t_{EQ} - t_i)/\tau$ decreases from
0.928 to 0.908. In any case over
90\% of the \BH life-time belongs to the matter dominated era. This is a 
consequence of the fact that the soft radiation produced at the early phase
of the \BH life-time further red-shifts during the expansion of the universe.
Using the expression \rf{realmatt} for cosmic scale parameter, one obtains
$(t_{EQ} - t_i)/\tau = 0.924$ which is in good agreement with the approximated
result. Therefore, it is reasonable to use the simpler equations. 

\begin{figure}[ht]
\leavevmode
\centering
\vspace*{60mm} 
\begin{picture}(0,60)(0,490)
\includegraphics{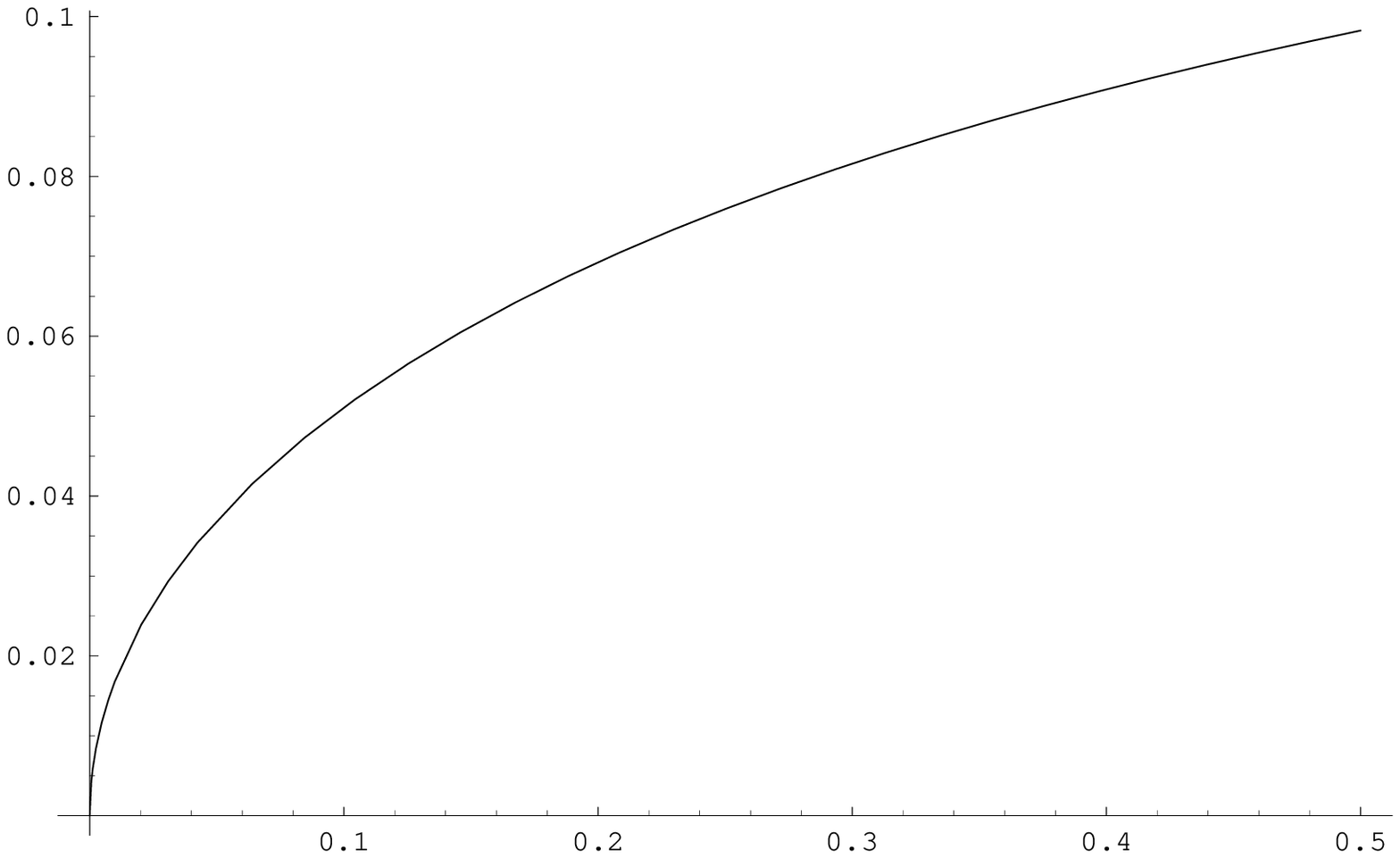}
  \put(-120,340){(a)}
\includegraphics{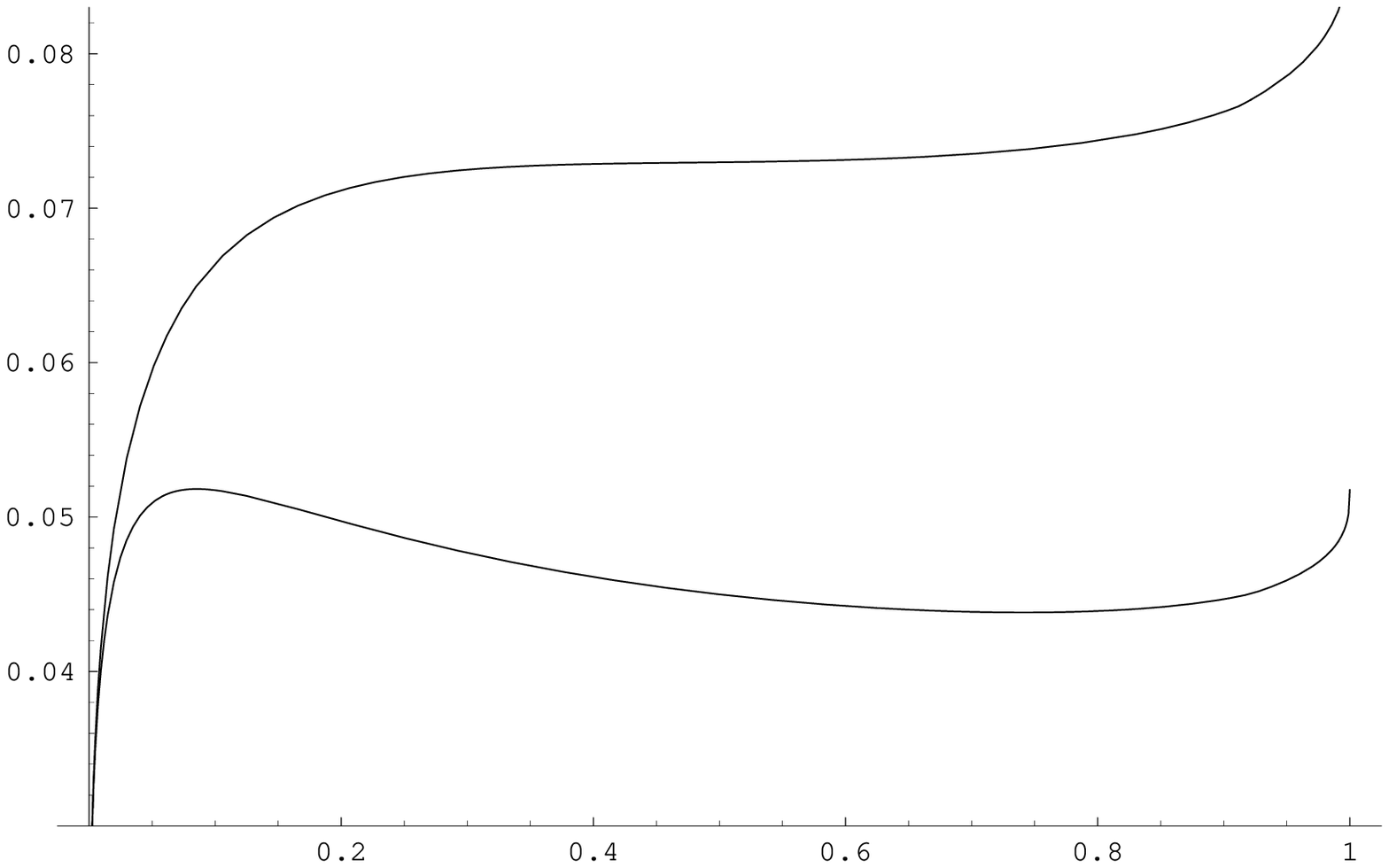}
  \put(100,340){(b)}
\end{picture}
\figcap{(a) The pseudo-temperature $\tilde T$ at $t = t_f$ as 
a function of the ratio $t_i/\tau$. ($t_i = t_{Pl}$, $\Omega_i = 1$.)
 (b) Evolution of the pseudo-temperature $\tilde T$ as a function of 
$(t - t_i)/\tau$. The upper curve corresponds $t_i/\tau = 0.35$
whereas the lower one $t_i/\tau = 0.1$. ($\Omega_i = 1$.)
\label{kuva1}
}       
\end{figure}

The scale factor is now simply given by equation
\be{scale}
R(t) =  \left [ \theta (t - t_{EQ}) \left ({t\over t_{EQ}}
\right )^{1/2}  + \theta(t_{EQ} - t) \left ({t\over t_{EQ}}\right )^{2/3} 
\right ] R(t_i)\left ({t_{EQ}\over t_i}\right )^{2/3},
\ee
where $\theta$ is the usual Heaviside-function. Eq. \rf{scale} can be inserted to formulas like
\rf{dis} and \rf{erad2}. In particular we are now interested in the energy density of the universe
at the end of the \BH evaporation era: the energy density at that time determines the later evolution 
of the universe.  We obtain the energy density at a given time $t$
\bea{etf}
\cE (t) &=& - \left({R(t_i)\over R(t_f)}\right )^3 n(t_i) \int_{t_i}^t  dt' \dot\mbh (t') {R(t')\over R(t_f)}
\nonumber \\
           &=& - \left ({t_i\over t_{EQ}}\right )^2 \left ({t_{EQ}\over t_f}\right )^{3/2} n(t_i)
\int_{t_i}^t dt' \dot\mbh (t') \nonumber \\
           & &\times \left [ \theta (t' - t_{EQ}) 
              \left ({t'\over t_{EQ}}\right )^{1/2}  +
              \theta(t_{EQ} - t') \left ({t_{EQ} \over t_f}\right )^{1/2} 
              \left ({t'\over t_{EQ}}\right )^{2/3} \right ],
\eea
where $t_i < t < t_f$. For times $t > t_f$ the energy density is only 
red-shifted as usual.
We define a pseudo-temperature $\tilde T$ to describe the
energy density and particle distribution by setting simply 
\be{tildeT}
{\pi^2\over 30} g_* \tilde T(t)^4 = \cE (t).
\ee
Now, $\tilde T$ is by no means a real temperature, simply because there exists 
no thermal equilibrium. However, it describes the energy density in familiar
way and, moreover, when the thermal equilibrium is finally maintained, 
$\tilde T$ coincides with the true temperature. Therefore, at the time 
$t_f$ it is enlightening to compare the actual particle distribution to 
the thermal one at pseudo-temperature $\tilde T$. In the Figure \ref{kuva1}a
we have displayed the pseudo-temperature $\tilde T$ at $t_f$ as a function of
$t_i/\tau$ ($t_i = t_{Pl}$). 

An interesting feature emerges in the behaviour of $\tilde T$ ({\it i.e.}
energy density). For $t_i/\tau < z_c \equiv 0.334$ the pseudo-temperature
has a minimum during the black hole evaporation era. In the Fig. \ref{kuva1}b 
we have plotted the evolution of two pseudo-temperatures, one with 
$t_i/\tau = 0.1 < z_c$ (lower curve) and another with $t_i/\tau = 0.35 > z_c$
(upper curve). We have scaled the time so, that $\tilde T$ is presented
as function of $ (t - t_i)/\tau$.
Note, that the large $t_i/\tau$ values correspond small \BHs with  
evaporation times comparable to $t_i$. Thus, if $t_i \simeq t_{Pl}$, the
approximations are not very good and. Practically in all relevant cases the 
pseudo-temperature has a minimum.

In this stage we have to make a notion about the homogeneity of the
matter and radiation in the universe. All calculations done as far, and 
all calculations to be done, assume that the universe is homogeneous at large 
scales. We can consider the matter to be homogeneously distributed
if there are several \BHs within a single horizon volume. On the other hand
the radiation can be viewed to be roughly homogeneous, if the distance between
the radiation sources, {\it i.e.} \BHs is shorter than the length propagated
by radiation. These two conditions are essentially equivalent leading to
$n(t) > t^{-3}$. In particular, this requirement at transition time $t_{EQ}$ 
means, that $\delta < 0.14 (\Omega_i)^{3/2}$. So, if $\Omega_i = 1$ we obtain
$\delta < 0.14$, which in terms of \BH mass reads 
$m_\bh  > 0.356 (h_*/160)^{1/3}$.

One can also ask, when the last emitted radiation (at $t = t_f$) is 
homogenized. Spatially homogeneous distribution is reached at time $t_h$ 
(assuming the radiation domination 
after $t_f$) if $n(t_f)^{-1/3} R(t_h)/R(t_f) < t_h - t_f$. Performing a 
numerical calculation it shows up (with $\Omega_i = 1$), that 
$(t_h - t_f)/\tau$ varies from 0 to 3 as $\delta$ increases from 0 to 0.5. Thus
the homogenization time is of the same order of magnitude as $t_f$ itself or
even shorter.

\begin{figure}[ht]
\leavevmode
\centering
\vspace*{55mm} 
\begin{picture}(0,60)(0,490)
\includegraphics{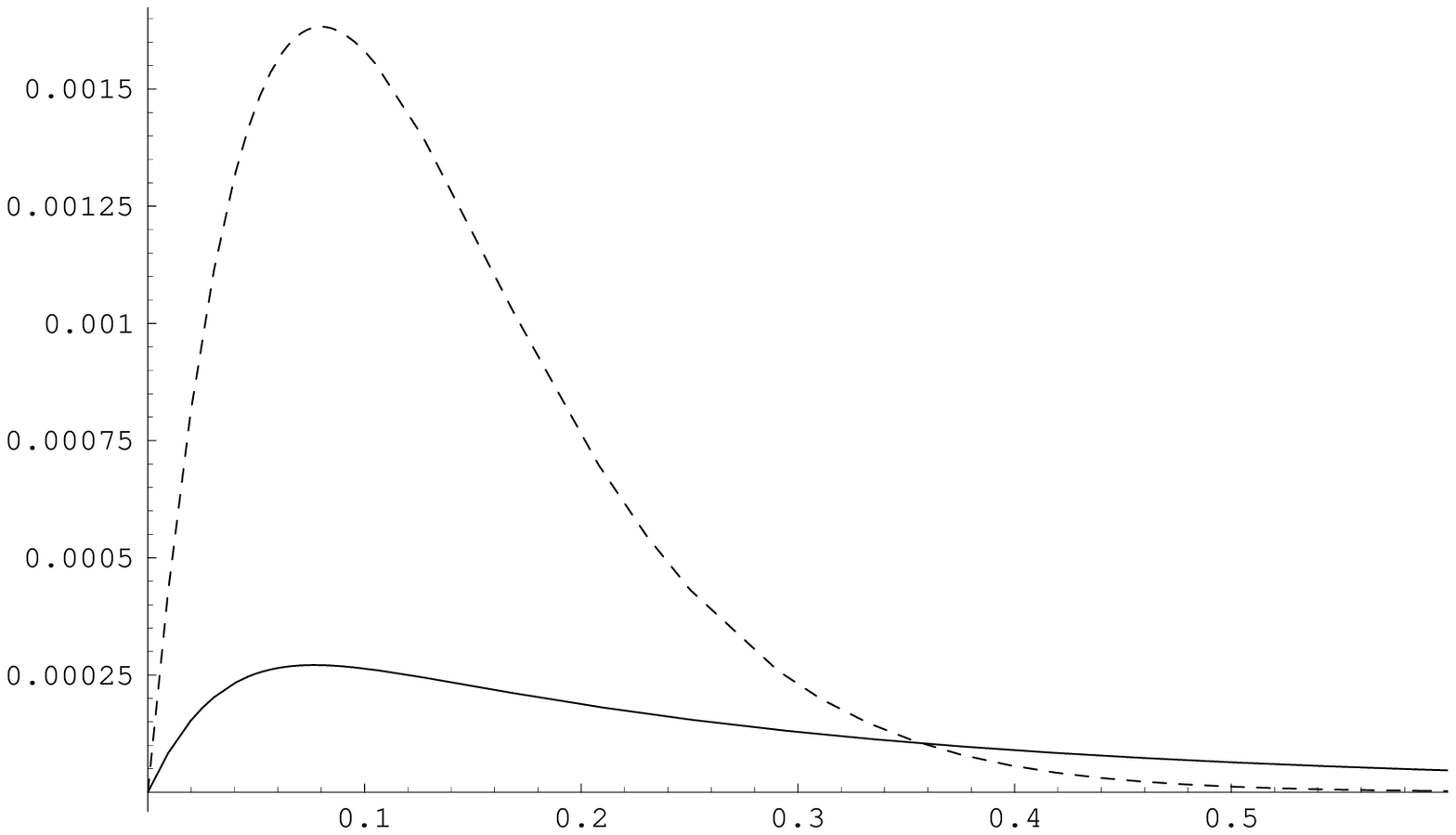}
  \put(-120,350){(a)}
\includegraphics{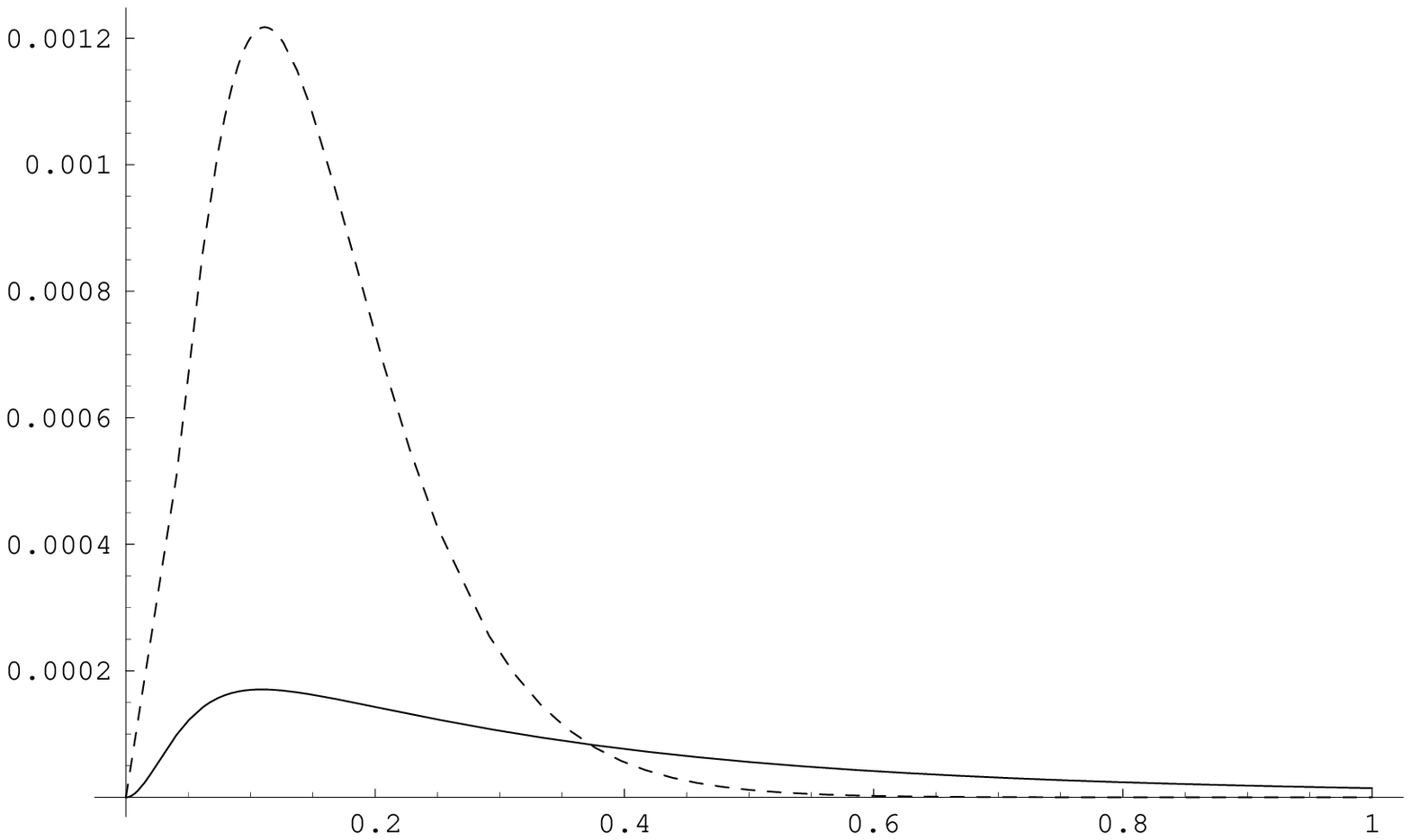}
  \put(100,350){(b)}
\end{picture}
\figcap{(a) Boson distribution function $p^2 f_b(p; t_f)$ as a function of
momentum $p$ (solid line). For reference, also the equilibrium distribution
at the temperature $\tilde T(t_f)$ has been presented (dashed line).
$\delta = 0.1$. (b) Corresponding fermion distribution function
$ p^2 f_f(p; t_f)$ (solid line) with equilibrium reference curve at $\tilde T(t_f)$ (dashed line). $\delta = 0.1$.
\label{kuva3}
}       
\end{figure}

Keeping in mind all above, we are now finally ready to calculate the 
distributions themselves. Applying Eqs. \rf{BHD}, \rf{sBHD}, \rf{scale}
and formulae for \BH density and temperature to the Eq. \rf{dis}, one obtains
the particle distributions. The distributions can be written as
\be{dis2}
f_{f(b)} (p;t) = {16 n_i\over M_{Pl}^4} \int_{t_i}^t dt'\, \mbh (t')^2
\left ({R(t_i)\over R(t')}\right )^3 f^\pm ({R(t)\over R(t')}
\beta_{BH}(t') p).
\ee
In the Fig. \ref{kuva3}a we have displayed the distribution at $t_f$ for bosons
multiplied by $p^2$ as a function of the momentum $p$ together with a 
reference distribution. The reference distribution is
the equilibrium distribution with $T = \tilde T(t_f)$. All  are given in Plank
units with $\Omega_i = 1$. The shape of distribution given in Fig. \rf{kuva3}a 
is generic and
it is noteworthy that the \BH originated distribution has  more power in 
large momenta. On the other hand, the IR end of the distribution, 
$p \sim 0$, there is less power.
Indeed, the IR behaviour of reads $f_b \sim \kappa/p$ like for 
equilibrium (where $f^- \sim T/p$), but 
\be{kappa}
\kappa = {2 n_i\over \pi M_{Pl}^2} {R(t_i)\over R(t_f)}
\int_{t_i}^{t_f} dt\, \mbh (t) \left ( {R(t_i)\over R(t)}\right )^2 
< \tilde T(t_f)
\ee
for all relevant $\delta$: the ratio
$\tilde T(t_f)/\kappa$ is for all $\delta < 0.5$ larger than $\sim 5.6$. 
This may be important, because the IR effects are essential in many physical 
considerations, in particular when some kind of effective potential and/or
interactions are used. 

The fermionic distribution function is presented in Fig. \ref{kuva3}b. 
As for bosonic case we have presented for comparison the equilibrium 
distribution at $\tilde T(t_f)$, too. The shape of the
distribution in Fig. \ref{kuva3}b is generic, and again there is more power at
large $p$ in $f_f$ than in equilibrium distribution. In general, the 
distributions arising from \BH evaporation are 
flatter and lower than corresponding equilibrium distributions. This can be
understood, because at late times $t \sim t_f$ the \BH radiates particles with 
large momenta. The particle density, $n_{b(f)}$, is thus always lower than 
at equilibrium with same energy density, $n_{b(f)}^{EQ}$. The ratio
$n_{b(f)}/n_{b(f)}^{EQ} < 0.36$ for all $\delta < 0.5$ and is proportional to
$\delta^{1/6}$ when $\delta \goto 0$.

Finally the question, how small the parameter $\delta$ can be, {\it i.e.}
how large the \BHs can be. If one supposes that the \BHs evaporate before the
thermalization takes place due to GUT processes \cite{1}, strong 
interactions \cite{ES} or electroweak interactions \cite{EEV1}, one has to
require that $t_f \ll t_{th} = 1/\Gamma$, where $\Gamma$ is the (average)
thermalization rate of the particles involved. If $\Gamma \sim \alpha \tilde T$,
where inspired by GUT's $\alpha \sim 3\times 10^{-2}$, we  conclude that 
$\delta/\tilde T \gg \alpha^2$ or $\delta \gg 3 \times 10^{-8}$. 
This estimate has many uncertainties
{\it e.g.} because as dimensional parameter the pseudo-temperature $\tilde T$ 
has been used. Nevertheless, this approximation gives a general idea about the
time-scales involved. On the other hand, one should address some interest to
the question, when the masses are negligible. Concerning
the tree-level masses, general remarks are impossible without specifying a
particular model. However, in the case of the ensemble masses ("thermal masses")
some remarks can be done. The ensemble mass $m_e \sim g \tilde T$, where
$g$ is a generic coupling, should be small compared to the \BH temperature 
$T_\bh$,
{\it i.e.} $m_e \ll T_\bh$. This can be rewritten as $\delta^{1/3}/\tilde T \gg
g \pi ({h_*\over 5\pi^2})^{1/3}M_{Pl}^{-1}$, where $T_\bh = T_\bh (t_i)$ has 
been used because it corresponds the lowest \BH temperature giving a 
conservative limit. Now the left hand side of this equation is always
larger than 7.9 (at $\delta = 0.5$). Thus (for $h_* = 160$) the ensemble mass 
can be ignored, if the coupling $g \ll 1.7$!

In the present, paper a general study of the particle distributions 
inspired by the \BH evaporation is presented. It is found that
the distribution produced differ clearly from the equilibrium distributions.
The distribution has much more power at large momenta and the boson 
distribution has slower increase  when $p\goto 0+$. The conditions of validity
of the analysis have been studied and found that under wide range of parameters
calculation can be performed with good reliability. Also the evolution of the 
universe has been studied during the \BH evaporation era. In particular, the 
time when it makes transition from matter domination to radiation domination 
and the generic bahaviour of radiation density are calculated.
These considerations may have great impact to the calculations of the very 
early universe. However, more detailed studies are definitely needed, when any 
particular model is considered.


\end{document}